\documentclass[reprint,aps,prl]{revtex4-1}
\usepackage[utf8]{inputenc}
\usepackage{amsmath}
\usepackage{amsfonts}
\usepackage{amssymb}
\usepackage{amsmath}
\usepackage{graphicx}
\usepackage{xcolor}
\usepackage{rotate}
\usepackage[outdir=./]{epstopdf}
\usepackage[english]{babel} % to be able using aps style bibliography
\usepackage{braket}
\usepackage{blindtext}
\usepackage{float}
\usepackage[left=2cm,right=2cm,top=2cm,bottom=2cm]{geometry}
\begin{document}
\title{Topological photonics: from crystals to particles}
\author{Gleb Siroki}
\author{Paloma A. Huidobro}	
\author{Vincenzo Giannini}
\affiliation{Department of Physics, Imperial College London}
\begin{abstract}
Photonic crystal topological insulators host protected states at their edges. In the band structure these edge states appear as continuous bands crossing the photonic band gap. They allow light to propagate unidirectionally and without scattering. In practice it is essential to make devices relying on these effects as miniature as possible. Here we study all-dielectric photonic topological insulator particles (finite crystals) which do not require a magnetic field. In such particles the edge states' frequencies are discrete. Nevertheless, the discrete states support pseudospin-dependent unidirectional propagation. They allow light to bend around sharp corners similarly to the continuous edge states and act as topologically protected whispering gallery modes which can store and filter light as well as manipulate its angular momentum. In addition, they explain multiple experimental observations of discrete transmission peaks in photonic topological insulators.
\end{abstract}
\maketitle 

\begin{figure}[h!]
\centering
\includegraphics[scale=0.5]{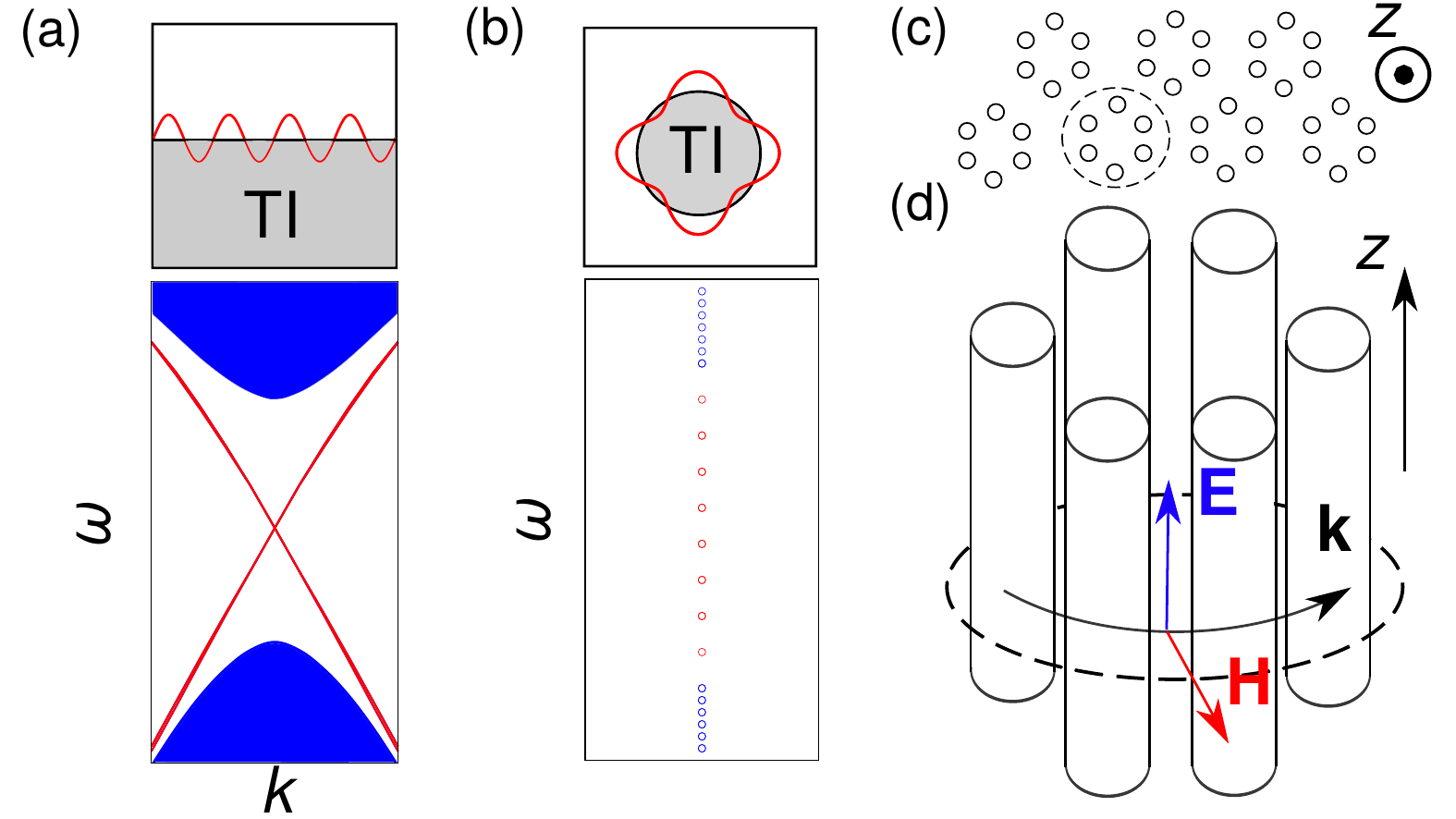}
\vspace{-3 mm}
\caption{(a) Top panel: an edge state of a photonic TI crystal. Bottom panel: the crystal band structure with blue representing the projected bulk bands. Red bands inside the bulk band gap are the topologically protected edge states. (b) Same for a TI particle -- red circles represent edge states. (c) Top view of the photonic TI made of dielectric rods used here. (d) Six rods constitute an artificial atom (encircled). \label{fig:1}}
\end{figure}

\textit{Introduction.} The invention of photonic crystals has greatly enriched the field of photonics \cite{yablonovitch_inhibited_1987,john_strong_1987}. More recently this field benefited from applying the ideas of topology to photonic band structures \cite{haldane_possible_2008}. This was triggered by the discovery \cite{kane_z_2_2005} and experimental confirmation \cite{konig_quantum_2007, hsieh_topological_2008} of an electronic topological insulator (TI). The hallmark of TIs are protected states occurring at the boundary of the crystal (Fig.~\ref{fig:1}(a)). These states support unidirectional propagation and are immune to certain defects. Their analogues are now being explored in acoustics \cite{prodan_topological_2009,khanikaev_topologically_2015,he_acoustic_2016}, optical lattices \cite{goldman_light-induced_2014}, plasmonics \cite{yuen-zhou_plexciton_2016, jin_infrared_2017,pan_topologically_2017,nalitov_polariton_2015} and especially photonics \cite{lu_topological_2014}. While photonic crystal analogues of 3D TIs have only been recently proposed \cite{lu_symmetry-protected_2016, slobozhanyuk_three-dimensional_2016}, their 2D counterparts have been already been realized experimentally \cite{lu_topological_2014, wang_observation_2009, rechtsman_photonic_2013, skirlo_experimental_2015,cheng_robust_2016, yang_visualization_2016}. Some of these realizations employ magnetic fields \cite{wang_observation_2009,skirlo_experimental_2015} (Quantum Hall effect) while others preserve the time-reversal symmetry \cite{khanikaev_photonic_2013,cheng_robust_2016, yang_visualization_2016} (Quantum Spin Hall effect). In the latter the angular momentum of light (spin) provides an additional degree of freedom \cite{shitrit_spin-optical_2013,hafezi_imaging_2013,dong_valley_2017}. The spin can be used within spin-chiral networks, spin-controlled gates and other devices for integrated optical circuits and quantum computing \cite{bliokh_spin-orbit_2015}. In contrast to conventional optical devices, these will be robust to manufacturing imperfections and able to take irregular shapes \cite{lu_topological_2016}. 
\newline
\indent Here we study a photonic TI particle (finite crystal). Because it is difficult to break time-reversal symmetry at optical frequencies \cite{lu_topological_2016}, the particle considered is made of dielectric rods \cite{wu_scheme_2015,ma_all-si_2016,milicevic_orbital_2017} as proposed by Wu and Hu \cite{wu_scheme_2015} and does not require a magnetic field. As opposed to a TI crystal (Fig.~\ref{fig:1}(a)) the particle has a discrete rather than continuous spectrum of edge states as shown in Fig.~\ref{fig:1}(b).
These states are a topologically protected version of whispering gallery modes \cite{yang_advances_2015}. They support unidirectional propagation determined by the pseudospin of light. Moreover, they exist for particles of different shapes and survive defects that do not strongly perturb the inherent crystal symmetry.
Finally, they explain the observation of discrete peaks in transmission \cite{wang_observation_2009, skirlo_experimental_2015, cheng_robust_2016, yang_visualization_2016} which occur because the photonic crystals studied in experiments are small and better described as particles.
\newline
\textit{Edge states of the photonic TI particle.} The parent photonic crystal studied is made from dielectric rods extending along the $z$-axis (Fig.~\ref{fig:1}(c)). Six of these rods constitute an artificial atom \cite{wu_scheme_2015} as shown in Fig.~\ref{fig:1}(d). The crystal made from these atoms arranged on a triangular lattice possesses $C_6$ and time-reversal symmetries. Their combination allows the crystal to become a Quantum Spin Hall insulator for TM polarization \cite{wu_scheme_2015}. %
Each atom carries four orbitals labelled $p_x,p_y,d_{x^2-y^2},d_{xy}$ according to their $E_z$ fields. The respective bands form a Dirac crossing at a planar interface of the photonic TI as illustrated in Fig.~\ref{fig:1}(a). A unitary transformation forms a new basis of $p_{\pm}=p_x \pm ip_y$ and $d_{\pm}=d_{x^2-y^2} \pm id_{xy}$. In this basis a photonic 4x4 $\bold k\cdot\bold p$ Hamiltonian can be found \cite{wu_scheme_2015} similar to the electronic Quantum Spin Hall Hamiltonian \cite{bernevig_quantum_2006} (see Supplemental Material). We apply the method of Imura \textit{et al.} \cite{imura_spherical_2012} to the TI particle with a circular cross section lying in the x-y plane.  The two solutions:
\begin{equation} \label{eq:minus}
\bold c_{+,m}=R(r)(e^{-i\phi},1,0,0)^T e^{im\phi}/\sqrt{2}
\end{equation}
\begin{equation} \label{eq:plus}
\bold c_{-,m}=R(r)(0,0,e^{i\phi},1)^T e^{im\phi}/\sqrt{2}
\end{equation}
are the coefficients in front of $p_+,d_+,p_-,d_-$ in the field expansion (Eq. 10 of Supplementary Material). The azimuthal number, $m=0, \pm 1, \pm 2...$, ensures periodicity, $c_{\pm,m}(2\pi+\phi)=c_{\pm,m}(\phi)$. The function $R(r)$ peaks near the particle's boundary. The states have frequencies
\begin{equation} \label{eq:dispersion}
w_{\pm,m}= w_0+\frac{C}{r_\text P}(\frac{1}{2}\pm m)
\end{equation}
where $w_0$ is in the middle of the band gap, $r_\text P$ is the particle's radius and $C$ -- some constant. Thus the states come in equally spaced pairs with $(+,m)$ and $(-,-m)$ being degenerate. A similar degeneracy exists in electronic TI nanoparticles due to time-reversal symmetry of the Schrodinger equation \cite{imura_spherical_2012, siroki_single-electron_2016}. However, photons are not fermions and require an additional symmetry to realize a TI. Here this is realized by the $C_6$ symmetry of the photonic crystal \cite{wu_scheme_2015} as expanded below.
\begin{figure}[h!]
\centering
\includegraphics[scale=0.4]{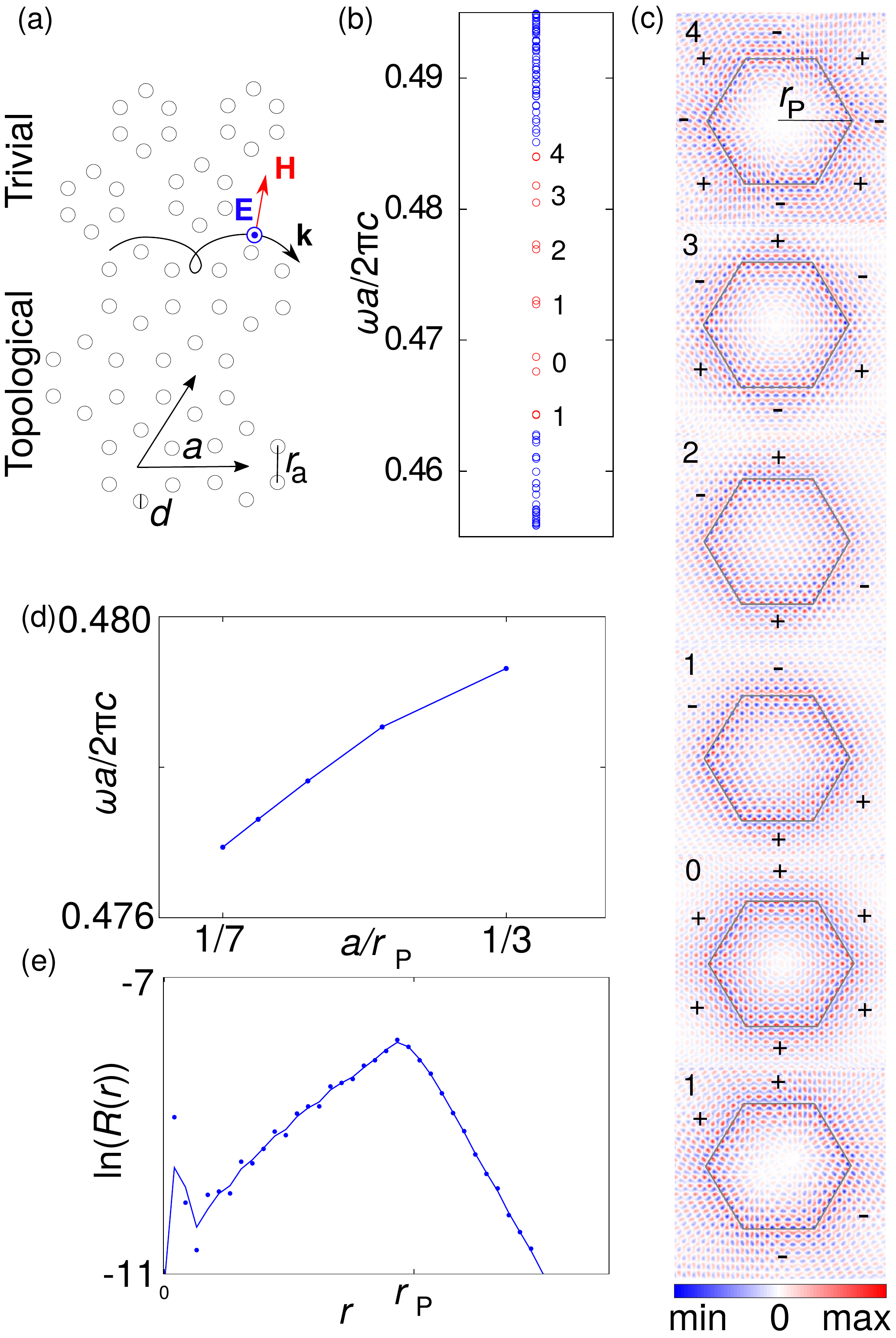}
\caption{(a) Interface between the topological and trivial photonic crystals supports protected edge states. (b) Frequency spectrum of a hexagonal photonic TI particle embedded in the trivial matrix ($\epsilon=11.7$). For the size considered (169 atoms) there are six pairs of TM edge states. They are labelled by the number of nodes in azimuthal direction. (c) $E_z$ fields for the states of the particle are shown (one state from each pair); $+$ and $-$ denote the relative phase of the field. (d) The frequency of the state labelled "$2$" decreases linearly with the particle's radius. This agrees with the analytical model -- see the main text. (e) The $E_z$ field averaged over azimuthal angles for the same state. The resulting radial distribution decays exponentially away from the edge. \label{fig:2}}
\end{figure}

To verify the analytical model we have performed numerical simulations for a hexagonal particle  %
(close to the circular shape). The constituent rods are made of silicon ($\epsilon=11.7$) with air in-between them as in \cite{wu_scheme_2015}. We simulate the hexagonal TI particle of radius, $r_\text P/a=7$ (169 atoms, $a$ -- separation between them). The particle is encased in a trivial matrix -- as in Fig.~\ref{fig:1}(b). The edge states appear at the boundary between the topological ($r_\text a> a/3$) and trivial atoms ($r_\text a< a/3$) as illustrated in Fig.~\ref{fig:2}(a). Their spectrum obtained with frequency domain simulations using the MPB package \cite{johnson_block-iterative_2001} is shown in Fig.~\ref{fig:2}(b). There are twelve states half of which is shown in Fig.~\ref{fig:2}(c). For the simulation details and the six $E_z$ fields not shown -- see the Supplementary Material. Each standing wave state is made of two counter-propagating waves. These two waves have opposite pseudospins ($H_x\pm i H_y$) required by the time-reversal symmetry of Maxwell's equations and each supports unidirectional propagation of light. The standing waves themselves come in pairs with regular spacing -- agreeing with Eq.~\ref{eq:dispersion}. The degeneracy is not exact because the particle is not circular. Also, the finite particle studied means the $C_6$ symmetry is only approximate due to the mismatch between the TI and trivial regions. The atoms have similar radii so the effect is weak (further decreasing with the size of the particle). In contrast, for electronic TI nanoparticles the necessary symmetry (time-reversal) is always present. There the states retain their degeneracy but get pushed out of the bulk gap as the size decreases \cite{imura_spherical_2012, siroki_single-electron_2016}. For the photonic TI particle Eq.~\ref{eq:dispersion} also implies that larger particles have more edge states. As the size increases, additional pairs of edge states emerge from the bulk bands. Their frequency scales with $1/r_\text P$ (Fig.~\ref{fig:2}(d)) in agreement with Eq.~\ref{eq:dispersion}. For very small sizes there is not enough bulk to support the topological states. This effect was also predicted for ribbons of the photonic TI considered \cite{wu_scheme_2015} and experimentally observed for slabs of electronic TIs \cite{zhang_crossover_2010}. It can be ascribed to the overlap of wave functions from opposite sides -- the states themselves decay exponentially away from the boundary as seen in Fig.~\ref{fig:2}(e). Finally, the numerical results agree with the analytical model in terms of the states' symmetries. To obtain a state that has no nodes in $E_z$ field we need to have the coefficients $e^{\mp i \phi}$ ($e^{\mp 2 i \phi}$) in front of $p_\pm$ ($d_\pm$) s.t. the field rotates as $p_x \cos \phi+p_y \sin \phi$ ($d_{x^2-y^2}\cos 2\phi +d_{xy}\sin 2\phi $). This ensures that the 'atomic' orbitals merely rotate as we go around the circumference and occurs for the states $(+,m=-2)$ and $(-,m=2)$. Both states have frequency $w=w_0-3C/2r_\text P$ according to Eq.~\ref{eq:dispersion}. The pair of states immediately above (below) have an additional factor of $e^{\pm i\phi}$ and should possess one node  which is indeed observed in Fig.~\ref{fig:2}(c). We stress that the degeneracy here is a result of the photonic crystal symmetry, c.f. electronic TIs. Each edge state (standing wave) is made of two time-reversed waves as required by Maxwell's equations. These edge states also occur for particles of other shapes. 
\begin{figure}[h!]
\centering
\includegraphics[scale=0.4]{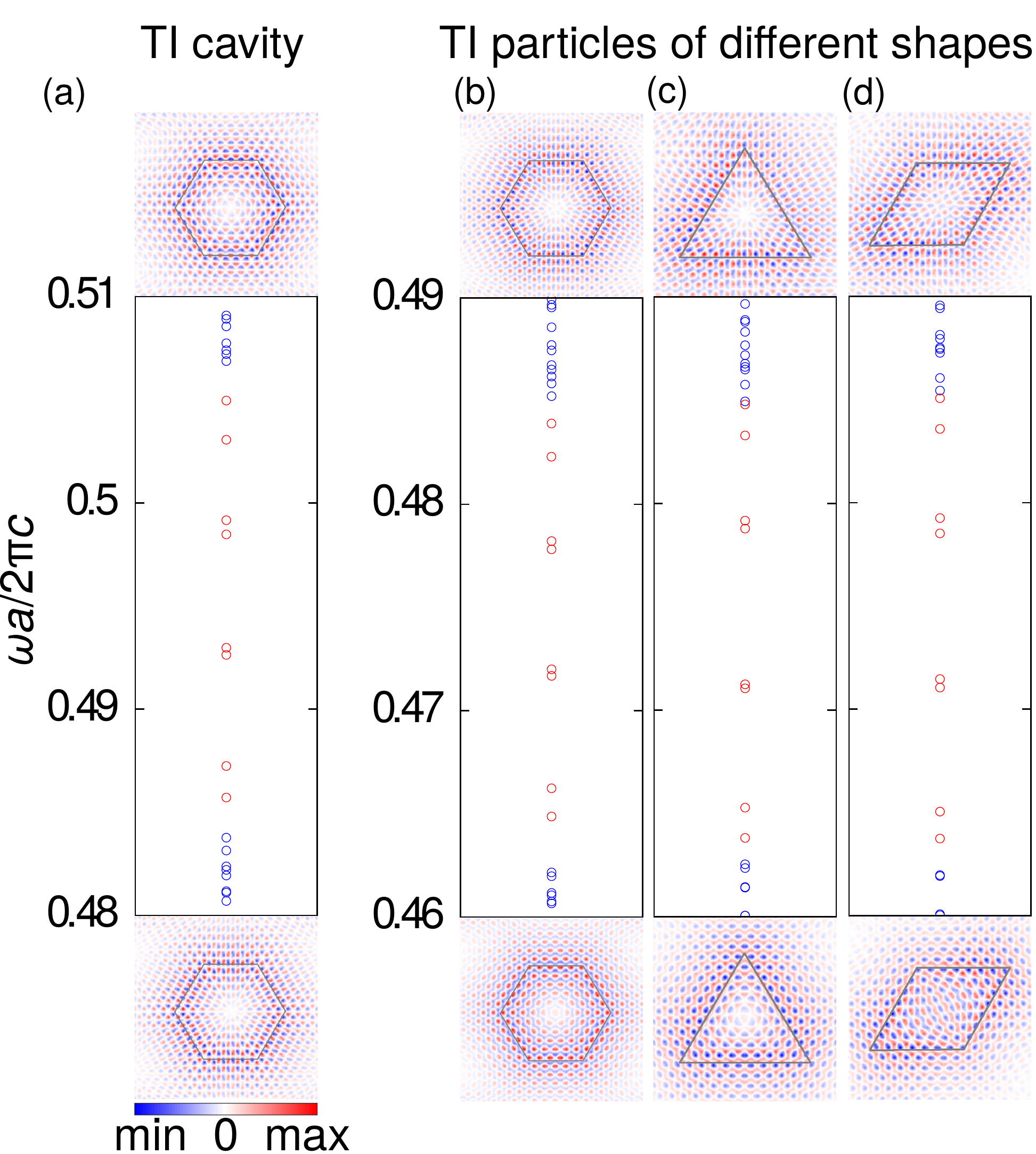}
\caption{The top (bottom) figure shows the $E_z$ field of the highest (lowest) edge state for each case considered ($\epsilon=11.7$): (a) The 'inverse' case of a trivial particle inside the topological matrix. For the size considered (91 atom) there are four pairs of TM edge states labelled by red circles. (b) A TI particle of the same size inside the trivial matrix. The field distributions are very similar and so are the spectra except for a rigid shift in frequency. (c, d) Triangular and rhombic TI particles (66 and 64 atoms). These cases show that the edge states exist for particles of different shapes and can bend around sharp corners. \label{fig:3}}
\end{figure} 

\begin{figure}[h!]
\centering
\includegraphics[scale=0.4]{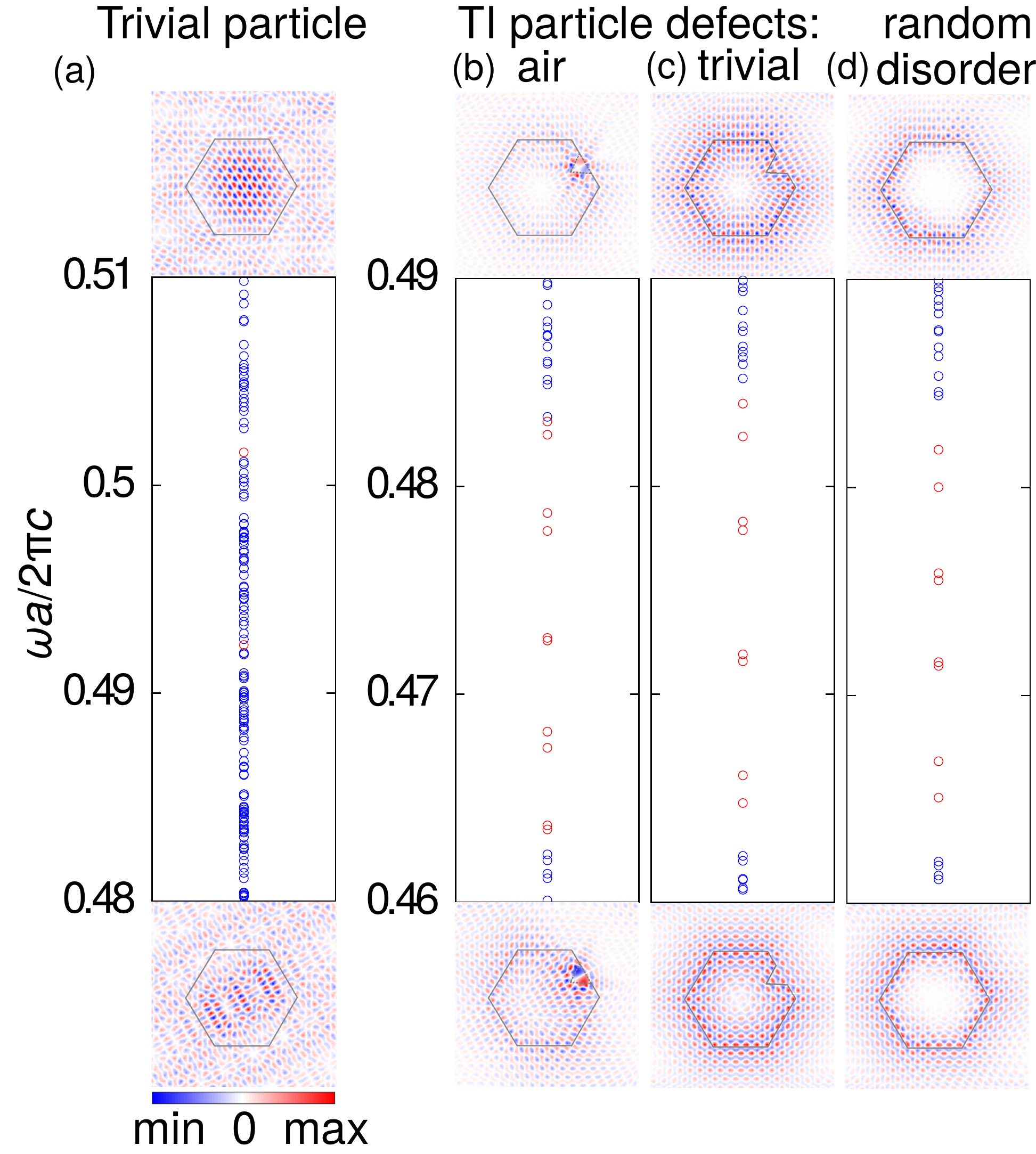}
\caption{The top (bottom) figure shows the $E_z$ field of the highest (lowest) state marked by a red circle for each particle considered (91 atoms, $\epsilon=11.7$): (a) A trivial particle inside the trivial matrix has no edge states. (b) A TI particle inside the trivial matrix with three atoms absent -- breaking the $C_6$ symmetry necessary for topological protection. This strongly affects the spectrum and localizes the edge states. (c) The same situation but with the three atoms inside the topological particle replaced by the trivial ones. This weakly affects the $C_6$ symmetry and hence -- the edge states. (d) A TI particle with positional disorder introduced. The coordinates of each rod in topological atoms are randomly changed by 10$\%$. This weakly perturbs the states suggesting they will survive in practice. \label{fig:4}}
\end{figure}

\indent \textit{Photonic TI particles of different shapes}. 
The edge states arise due to bulk band structure and should be present for particles of any shape that are large enough and preserve relevant symmetries. Before proceeding to investigate the influence of shape however we consider a cavity,i.e. a trivial hexagonal particle inside the topological matrix. %
The analytical solution here is the same except the $R(r)$ part which now extends outside from the cavity (see Supplementary Material). This agrees with Fig.~\ref{fig:3}(a) where the cavity states are practically identical to those of the particle. The spectrum is also very similar apart from a rigid shift in the frequencies of the states -- Fig.~\ref{fig:3}(b) shows the hexagonal particle with the same number of atoms ($a/r_\text P=5$, 91 atom). Note how the degeneracy of the states is worse than in Fig.~\ref{fig:2}(c) where the bigger particle has 'more' of the $C_6$ symmetry. To explore the shape effects we have computed the spectra for the particles of triangular and rhombic shapes (66 and 64 atoms) -- as presented in Fig.~\ref{fig:3}(c) and (d). These contrast sharply  with the case of a trivial particle inside a trivial matrix shown in Fig.~\ref{fig:4}(a). The TI particles have the states that occur in pairs and are able to bend around sharp corners. The latter feature has also been observed experimentally \cite{yang_visualization_2016}. This insensitivity to the shape confirms the topological origin. In addition, such states should also be immune against certain defects.

\indent \textit{Photonic TI particle with a defect.} The edge states of electronic topological insulators are protected against defects and impurities which are non-magnetic. The latter is important because magnetic disorder breaks the required (time-reversal) symmetry. For the photonic TI crystal here this can be achieved by breaking the $C_6$ symmetry of the lattice \cite{lu_topological_2016}. A strong symmetry breaking occurs when we completely remove three atoms from the TI particle as shown in Fig.~\ref{fig:4}(b). As expected, this modifies the spectrum and localizes the edge states. In contrast, replacing the topological atoms with trivial ones is a relatively weak perturbation of the $C_6$ symmetry which hardly affects the spectrum and the edge states as seen in Fig.~\ref{fig:4}(c). This suggests that the states survive weak breaking of the $C_6$ crystal symmetry such as due to slightly different radii of the atoms in the TI particle and the trivial matrix. Finally, in practice the constituent rods can be displaced from their ideal positions. We have simulated this case by displacing the rods in each topological atom randomly by 10$\%$. The results in Fig.~\ref{fig:4}(d) show that the states are weakly affected. They suggest that the states can be realized and will support unidirectional propagation.

\begin{figure}[H]
\centering\includegraphics[scale=0.4]{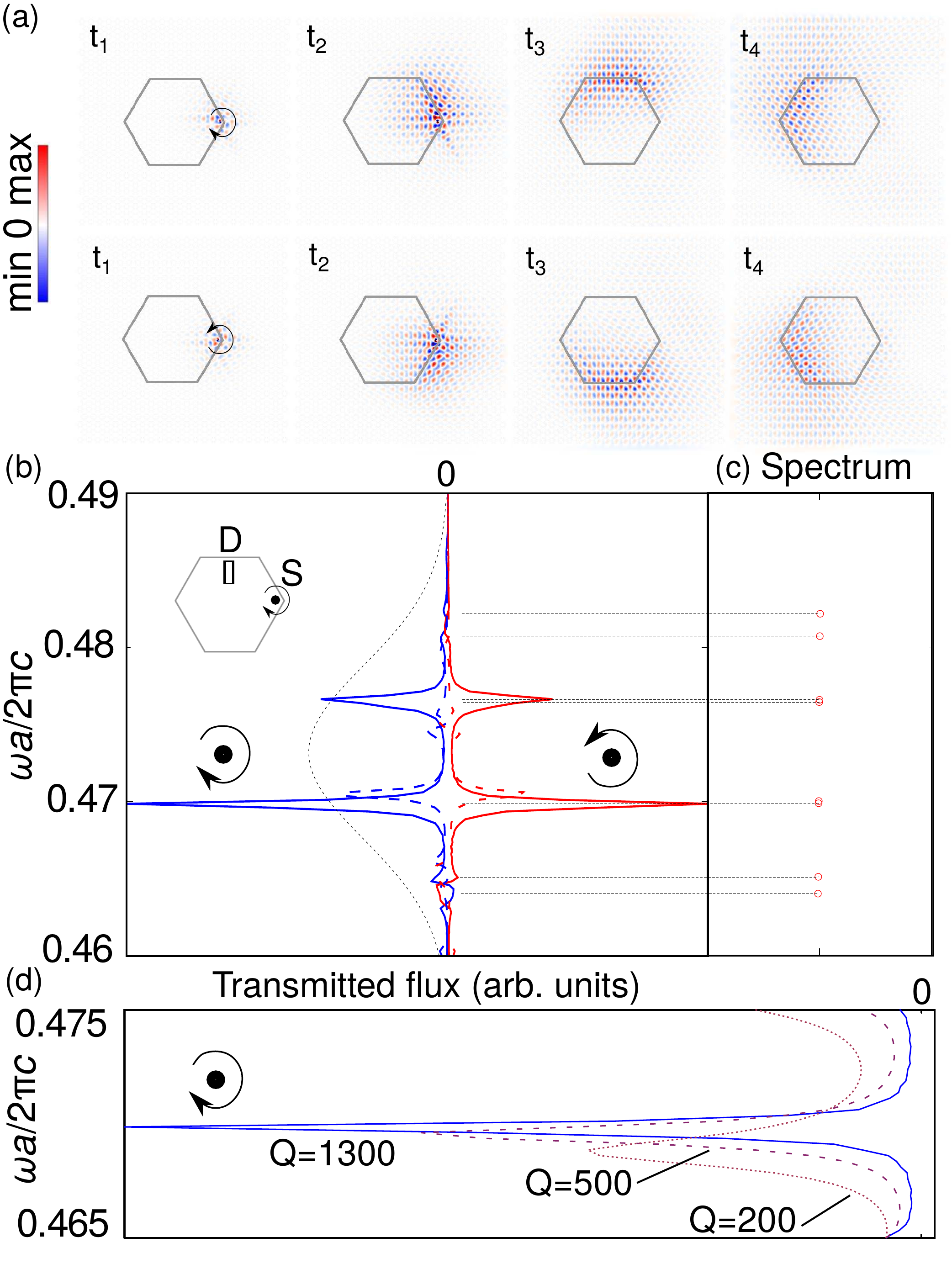}
\caption{Time-domain simulations of the photonic TI particle (91 atoms, $\epsilon=11.7$) edge states. (a) $E_z$ field of a propagating Gaussian pulse occurs in opposite directions depending on its pseudospin. (b) Inset: the particle, source (S) and detector (D). The transmission spectrum shows a number of discrete peaks -- positive values correspond to transmission in x-direction. Blue (red) solid lines correspond to $H_x+iH_y$ ($H_x-iH_y$) pseudospin. Dashed lines show corresponding results with 10$\%$ disorder in positions of rods constituting the particle (see the main text). Black dotted line shows the profile of the input Gaussian pulse (not to scale). (c) Horizontal lines show the frequencies of the  states obtained with time-domain simulations. (d) The width of the states depends on the thickness of the surrounding trivial matrix. Making the matrix 2 (4) trivial atoms thinner than above decreases the largest quality factor down to 500 (200) as indicated by the dashed (dotted) line. \label{fig:5}}
\end{figure}

\textit{Unidirectional propagation of light}. In electronic TIs the time-reversal symmetry restricts the edge states with opposite spins to travel in opposite directions. Analogously, in the photonic TI considered TM modes can have opposite pseudospins ($H_x\pm iH_y$). For a slab it has been shown \cite{wu_scheme_2015} that continuous edge states inside the band gap support propagation in either one or the other direction depending on the pseudospin. To show this for the discrete states in the particle we used the MEEP finite-difference time-domain package \cite{oskooi_meep:_2010}. The TI particle is surrounded by the trivial matrix and the simulation cell walls are covered with a perfectly matched layer (see Supplementary Material). We excite the edge states with two TM-polarized dipoles 90$^o$ out of phase with each other ($H_x \pm iH_y$). They emit a Gaussian pulse whose pseudospin determines its direction of propagation according to Fig.~\ref{fig:5}(a). The power transmission through a detector (normalized by the Q-factor) in Fig.~\ref{fig:5}(b) shows unidirectional propagation.
The transmission spectrum confirms that each edge state (standing wave) consists of two counter-propagating waves as required by the time-reversal symmetry of Maxwell's equations. Either one or the other of these waves is excited depending on the pseudospin of the incident light. In addition, we considered transmission for the particle with positional disorder as previously in Fig.~\ref{fig:4}(d). The resultant spectrum still shows unidirectional transmission for each given pseudospin as seen in Fig.~\ref{fig:5}(b). MEEP also allows to obtain complex frequencies of the edge states. Their real parts shown in Fig.~\ref{fig:5}(c) agree well with the frequency domain simulations. The quality factor, $Q=O(10^3)$, already for the trivial coating that is several atoms thick. Thinner layers lead to more leakage as well as $C_6$ symmetry breaking and hence backscattering as is evident from Fig.~\ref{fig:5}(d). Finally, we note that the unidirectional propagation for the photonic TI considered has been recently observed \cite{yang_visualization_2016}. In fact, the sample used in \cite{yang_visualization_2016} is comparable in size to the TI particles studied here. Therefore it should possess a discrete rather than continuous spectrum of edge states. We believe that this is the reason why discrete transmission peaks have been observed. \newline
\indent \textit{Conclusions}. To summarize, we have investigated for the first time discrete edge states that emerge in a photonic TI particle. This particle made of dielectric rods is a particular realization of a TI which does not require magnetic fields. The particle's spectrum contains edge states that occur in pairs. The spacing between the pairs decreases with the particle's size tending to the continuum limit. Each edge state, in turn, is made up of two counter-propagating waves with opposite pseudospins. These edge states are insensitive to the particle's shape as explicitly illustrated for hexagonal, rhombic and triangular particles. They are robust against certain defects and disappear only for very small particles. The states support propagation in one direction along the edge which can be switched with pseudospin. In practice, such microcavity can be used to manipulate photons. We expect interesting effects in arrays of such particles. Moreover, the edge states of TI particles will have peculiar effect on the photonic local density of states and radiation of chiral molecules.

\bibliographystyle{apsrev4-1}
%\bibliography{../../thesis}

\end{document}